\documentclass[
  aip,
  reprint,
  superscriptaddress
]{revtex4-1}

\usepackage{graphicx}
\graphicspath{{./figures/}} 
\usepackage{dcolumn}
\usepackage{bm}
\usepackage{hyperref}
\usepackage{physics} 
\usepackage{color}
\usepackage{empheq}

\begin{document}
\newcommand{\eqname}{Eq.}
\newcommand{\secname}{Sec.}

\title{A mechanism for anomalous transport in chiral active liquids}

\author{Zhenghan Liao}
\affiliation{Department of Chemistry, University of Chicago, Chicago, IL, 60637, USA}

\author{Ming Han}
\affiliation{James Franck Institute, University of Chicago, Chicago, IL, 60637, USA}
\affiliation{Pritzker School of Molecular Engineering, University of Chicago, Chicago, IL, 60637, USA}

\author{Michel Fruchart}
\affiliation{James Franck Institute, University of Chicago, Chicago, IL, 60637, USA}
\affiliation{Department of Physics, University of Chicago, Chicago, IL, 60637, USA}

\author{Vincenzo Vitelli}
\affiliation{James Franck Institute, University of Chicago, Chicago, IL, 60637, USA}
\affiliation{Department of Physics, University of Chicago, Chicago, IL, 60637, USA}

\author{Suriyanarayanan Vaikuntanathan}
\email{svaikunt@uchicago.edu}
\affiliation{Department of Chemistry, University of Chicago, Chicago, IL, 60637, USA}
\affiliation{James Franck Institute, University of Chicago, Chicago, IL, 60637, USA}

\date{\today}

\begin{abstract}
Chiral active fluids are known to have anomalous transport properties such as the so-called odd viscosity. In this paper, we provide a microscopic mechanism for how such anomalous transport coefficients can emerge. We construct an Irving-Kirkwood-type stress tensor for chiral liquids and express the transport coefficients in terms of orientation-averaged intermolecular forces and distortions of the pair correlation function induced by a flow field. We then show how anomalous transport properties can be expected naturally due to the presence of a transverse component in the orientation-averaged intermolecular forces and anomalous distortion modes of the pair correlation function between chiral active particles. We anticipate that our work can provide a microscopic framework to explain the transport properties of non-equilibrium chiral systems. 
\end{abstract}


\maketitle

\section{Introduction}

Recent work on active and driven matter systems have significantly advanced our understanding of how non-equilibrium forces can be used to modulate properties of soft matter systems and materials \cite{Marchetti2013HydrodynamicsSoft,Marchetti2016MinimalModel,Bechinger2016ActiveParticles,Ramaswamy2017ActiveMatter,Martinez2017ColloidalHeat}. 
In this paper, we focus on a particular class of active matter systems, namely, chiral active liquids \cite{Furthauer2012ActiveChiral,Banerjee2017OddViscosity,Julicher2018HydrodynamicTheory}. In these systems a non-equilibrium steady state is sustained through a steady injection of energy into the rotational degrees of freedom of each constituent molecule or rotor. 
Studies of such chiral active fluids in recent years have discovered a variety of interesting phenomena, including rich phase behavior \cite{vanZuiden2016SpatiotemporalOrder,Scholz2018RotatingRobots}, non-equilibrium self-assembly \cite{Snezhko2016ComplexCollective,Yan2016ReconfiguringActive,Aubret2018TargetedAssembly}, ordered phase stabilization \cite{Maitra2019SpontaneousRotation}, and localized mass currents at a boundary or an interface \cite{Tsai2005ChiralGranular,Nguyen2014EmergentCollective,vanZuiden2016SpatiotemporalOrder}.
Importantly, various studies have shown how the transport properties of chiral active systems are influenced by anomalous coefficients such as the so-called ``odd" or Hall viscosity~\cite{Banerjee2017OddViscosity,Souslov2019TopologicalWaves,Soni2018FreeSurface}.  
Unlike the conventional shear viscosity which leads to a resistance parallel to the direction of a velocity gradient, the odd/Hall viscosity is responsible for an anomalous response in the transverse direction.
Odd viscosity was postulated to be a generic feature of parity symmetry-breaking fluids \cite{Avron1998OddViscosity}, and it has been studied for systems like gases under Lorentz-like forces \cite{Chapman1990MathematicalTheory,Souslov2019TopologicalWaves}.
However, until recently \cite{Banerjee2017OddViscosity}, this term has not typically been included in hydrodynamic descriptions of chiral active fluids, possibly because its importance was not clear in many previously studied systems \cite{Tsai2005ChiralGranular,Furthauer2012ActiveChiral,vanZuiden2016SpatiotemporalOrder}. 
An understanding of odd viscosity from a molecular perspective, in particular the role played by activity, will be beneficial and aid in building a connection between the microscopic structure and the macroscopic transport properties of active chiral liquids \cite{Klymko2017StatisticalMechanics,Epstein2019TimeReversal}. 

In this paper, we provide a mechanism for how odd viscosity and other anomalous transport coefficients can emerge due to non-equilibrium activity in chiral active liquids. 
Our approach is motivated by Irving and Kirkwood's seminal work on the statistical mechanical theory of transport properties \cite{Irving1950StatisticalMechanical,Kirkwood1946StatisticalMechanical,Kirkwood1949StatisticalMechanical}. Specifically, we adapt Irving and Kirkwood's techniques\cite{Dahler1959TransportPhenomena,Evans1976GeneralizedHydrodynamics,Evans1978TransportProperties,Klymko2017StatisticalMechanics} and show that the stress tensor of a chiral liquid can be expressed in terms of orientation-averaged intermolecular forces and pair correlation functions computed using the center of mass locations of the chiral active molecules. Our central results then show how anomalous transport coefficients can naturally emerge due to the presence of a transverse component in the orientation-averaged force and anomalous modes in the distortion of the pair correlation function induced by flows. Features such as a transverse component in the orientation-averaged force can be sustained due to the non-equilibrium activity in the chiral liquid.

The rest of this paper is organized as follows.
In \secname~\ref{sec:stress}, we review the Irving-Kirkwood method and adapt this technique to compute the stress tensor for chiral active fluids.
In \secname~\ref{sec:extract}, we extract transport coefficients from the stress tensor, and show how anomalous transport coefficients can naturally emerge in chiral active systems. Central theoretical results for transport coefficients are formulated in \eqname~\eqref{eq:stress_terms}-\eqref{eq:stress_odd_bulk}. 
In \secname~\ref{sec:simulation}, we describe results from numerical simulations of an active rotor systems that support our theoretical predictions.

\section{Stress tensor for chiral active fluids} \label{sec:stress}

\begin{figure}[tbp]
	\centering
	\includegraphics[width=0.48\textwidth]{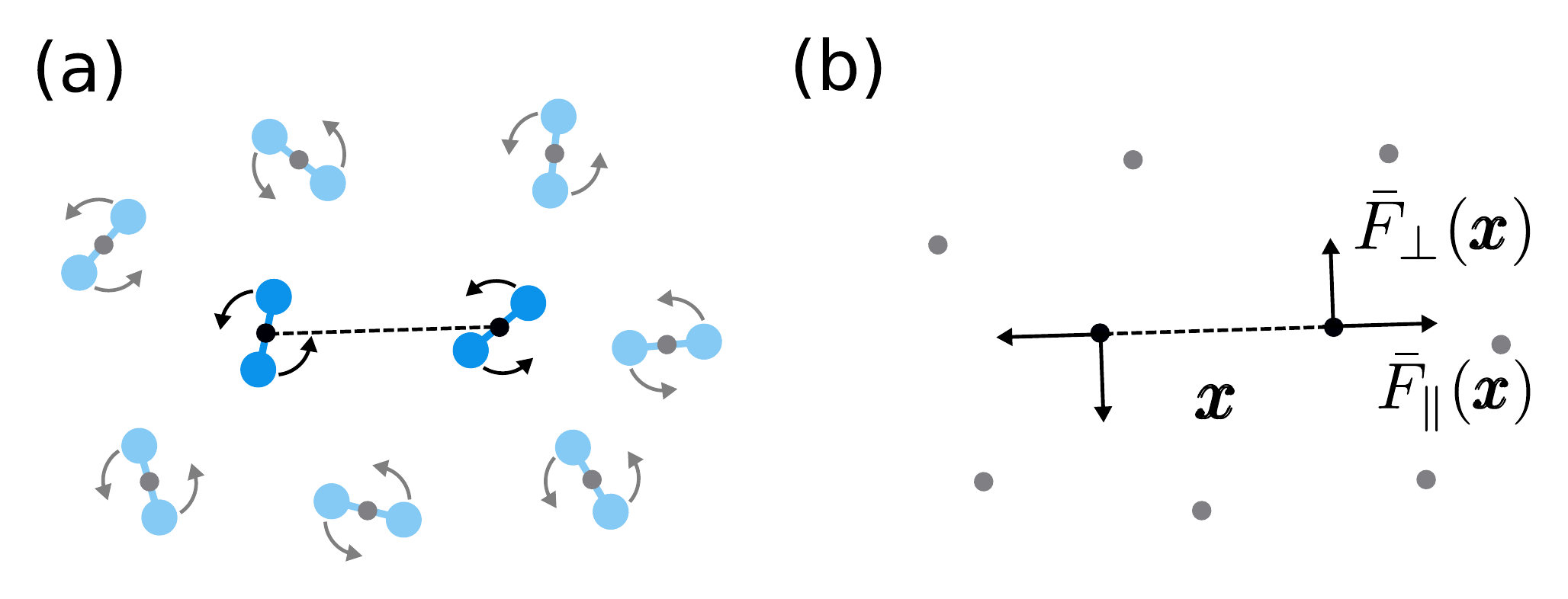}
    \caption{
        Schematic of the type of system we consider and the emergence of perpendicular averaged force.
        (a) A chiral active system with interacting rotors experiencing active torques.
        (b) After averaging over the orientation of rotors, the orientation-averaged intermolecular forces generically contain both an ordinary radial component $\overline{\bm{F}}_\parallel(\bm{x})$ and a peculiar perpendicular component $\overline{\bm{F}}_\perp(\bm{x})$. We will show that the anomalous transport coefficients can be expressed in terms of this perpendicular force.
    }
    \label{fig:schem_force}
\end{figure}

In this section, we review and adapt the Irving-Kirkwood technique~\cite{Irving1950StatisticalMechanical} to construct a stress tensor for chiral active liquids. Irving-Kirkwood-type stress tensors generally consist of a kinetic part and a potential part. In this work we only focus on the potential part, which is the dominant contribution for liquids \cite{Irving1950StatisticalMechanical,Brush1962TheoriesLiquid}. We do not consider the kinetic part, which becomes important for dilute gas-like systems~\cite{Irving1950StatisticalMechanical,Brush1962TheoriesLiquid}.
The potential part of the stress tensor will be expressed in terms of the orientation-averaged intermolecular forces and the pair correlation function of the center of mass of each molecule. We will show in subsequent sections how these two factors can be modified due to chiral activity, and how the modifications combine to produce anomalous transport properties.
The chiral active liquids are assumed to be two-dimensional ($2$D) or quasi-$2$D, single-component and homogeneous. 

The starting point for our derivation is the mechanical definition of the stress tensor $\bm{\sigma}$ as $\bm{\sigma}\cdot\dd{\bm{S}} = \text{``force across } \dd{\bm{S}} \text{''}$, where $\dd{\bm{S}}$ is a line element vector pointing from the inside of the line to the outside. 
Here, ``Across'' means that the vector $\bm{x}$, which connects the center of mass of a molecule inside of $\dd{\bm{S}}$ to one outside, intersects with $\dd{\bm{S}}$ at position $\bm{X}$.
The force across $\dd{\bm{S}}$ acting on the inside of $\dd{\bm{S}}$ can be written as
\begin{equation} \label{eq:stress_def}
    \begin{split}
        \bm{\sigma} \cdot \dd{\bm{S}}
        &= -\int\dd{\bm{\theta}}\int\limits_{\bm{x}\cdot\dd{\bm{S}}>0}\dd{\bm{x}}\int_0^1\dd{\alpha}
        \bm{F}(\bm{x};\bm{\theta}) \\ 
        &\rho^{(2)}(\bm{X}-\alpha\bm{x}, \bm{X}+(1-\alpha)\bm{x}; \bm{\theta}) \bm{x}\cdot\dd{\bm{S}}.
    \end{split}
\end{equation}
In this expression, $\bm{\theta} = \qty{\theta_1, \theta_2}$ denotes the orientation of the two molecules with respect to fixed axis, $\bm{F}(\bm{x};\bm{\theta})$ is the intermolecular force, $\rho^{(2)}(\bm{X}-\alpha\bm{x}, \bm{X}+(1-\alpha)\bm{x})$ is the two-body density, and 
$\rho^{(2)}(\bm{X}-\alpha\bm{x}, \bm{X}+(1-\alpha)\bm{x}; \bm{\theta}) \dd{\bm{S}}\cdot\bm{x} \dd{\alpha} \dd{\bm{x}}$ 
describes the probability of finding one molecule around $\bm{X}-\alpha\bm{x}$ with orientation $\theta_1$ and another around 
$\bm{X}+(1-\alpha)\bm{x}$ with orientation $\theta_2$, where $\alpha \in [0, 1]$ is a parametrization of the location of the molecule pair.
From the expression for $\bm{\sigma} \cdot \dd{\bm{S}}$, the stress tensor $\bm{\sigma}$ can be identified.

The expression for the stress tensor can be simplified using the homogeneity of the system.
Due to homogeneity or translational invariance, the two-body density $\rho^{(2)}(\bm{X}-\alpha\bm{x}, \bm{X}+(1-\alpha)\bm{x}; \bm{\theta})$ reduces to $\rho^{(2)}(\bm{x}; \bm{\theta})$, where the relative positions of the two molecules $\bm{x}$ replaced their absolute positions $\bm{X}-\alpha\bm{x}, \bm{X}+(1-\alpha)\bm{x}$.
Next we express the two-body density in terms of conditional probabilities and the pair correlation function,
\begin{equation} \label{eq:rho2_simp}
    \rho^{(2)}(\bm{x}; \bm{\theta})
    = p(\bm{\theta} | \bm{x}) \rho^{(2)}(\bm{x})
    = p(\bm{\theta} | \bm{x}) \rho^2 g(\bm{x}).       
\end{equation}
Here $p(\bm{\theta} | \bm{x})$ denotes the probability density of the orientation of the molecule pair conditioned on their positions. $\rho^{(2)}(\bm{x})$ denotes the two-body density regardless of molecular orientations, which equals to the product of density $\rho$ squared and the pair correlation function regardless of orientations, $g(\bm{x})$.
Plugging \eqname~\eqref{eq:rho2_simp} into \eqname~\eqref{eq:stress_def}, the stress tensor becomes
\begin{align}
    \bm{\sigma}
    &= -\int\dd{\bm{x}}\dd{\bm{\theta}} \frac{1}{2} \bm{F}(\bm{x};\bm{\theta})\bm{x} p(\bm{\theta} | \bm{x}) \rho^2 g(\bm{x}),
\end{align}
where the integral over the whole space $\int\dd{\bm{x}}/2$ is converted from $\int_{\bm{x}\cdot\dd{\bm{S}}>0}\dd{\bm{x}}$. This expression can be further simplified by defining an orientation-averaged intermolecular force, 
\begin{equation} \label{eq:force_avg}
    \bar{\bm{F}}(\bm{x}) = \int\dd{\bm{\theta}}\bm{F}(\bm{x};\bm{\theta}) p(\bm{\theta} | \bm{x}).
\end{equation}
The Irving-Kirkwood-type stress tensor now reads
\begin{align} \label{eq:stress_IK}
    \bm{\sigma} &= -\int\dd{\bm{x}} \frac{1}{2}\bar{\bm{F}}(\bm{x})\bm{x} \rho^2 g(\bm{x}).
\end{align}
Starting with \eqname~\ref{eq:stress_IK}, we demonstrate in \secname~\ref{sec:extract} how expressions for various transport coefficients can be extracted. 

Under conditions of equilibrium, the orientation-averaged force between two molecules in a chiral fluid will only contain a radial component. 
For chiral fluids out of equilibrium, however, the averaged force can contain a component that is perpendicular to $\bm{x}$ (sketched in \figurename~\ref{fig:schem_force}). We denote this perpendicular force as $\bar{\bm{F}}_\perp(\bm{x})$. 
The emergence of $\bar{\bm{F}}_\perp(\bm{x})$ can be motivated by a thought experiment using a system composed of a dilute gas of rotors interacting according to a repulsive potential (\figurename~\ref{fig:schem_force}). 
At equilibrium, the system obeys parity symmetry, which constrains $\bar{\bm{F}}_\perp(\bm{x})$ between rotors to be zero.
When driven out of equilibrium by driving each rotor with an active torque, the system violates parity symmetry, which can allow for a nonzero $\bar{\bm{F}}_\perp(\bm{x})$. Mechanistically, when the arms of the two rotors rotate toward each other, rotors' rotation would be hindered due to repulsive interactions, thus more time would be spent on this configuration. As a result, this configuration gains more statistical weight and biases $\bar{\bm{F}}_\perp(\bm{x})$ towards a nonzero value. This simple argument becomes intractable for dense liquid systems with many interacting rotors. In \secname~\ref{sec:sim_force}, we use numerical simulations to study how a transverse component in the orientation-averaged force can emerge due to non-equilibrium activity.  
 
We note debates in literature concerning the uniqueness of the Irving-Kirkwood stress tensor \cite{Irving1950StatisticalMechanical,SchofieldP.1982StatisticalMechanics,Wajnryb1995UniquenessMicroscopic,Goldhirsch2010StressStress,Admal2010UnifiedInterpretation}. However, most of these issues emerge due to the presence of inhomogeneities, such as due to the presence of surfaces or interfaces and are not important for the kind of homogeneous systems considered in this work.

\section{Extracting transport coefficients from the stress tensor} \label{sec:extract}

Transport coefficients are linear response coefficients of the stress tensor under a velocity gradient. To extract transport coefficients from the stress tensor in \eqname~\eqref{eq:stress_IK}, we investigate how \eqname~\eqref{eq:stress_IK} changes in the presence of a flow field using a method motivated by Ref.~\cite{Kirkwood1949StatisticalMechanical,Avron1998OddViscosity}.
Upon an application of such a velocity gradient or flow field, the stress tensor can respond due to a change in either the pair correlation function $g(\bm{x})$, or the orientation-averaged force $\bar{\bm{F}}(\bm{x})$. 
In this work, we only consider the distortion of the pair correlation function induced by the flow field in order to extract the transport coefficients \cite{Kirkwood1949StatisticalMechanical,Clark1980ObservationCoupling}.
Consequently, the expressions we derive for the various transport coefficients are only accurate when the orientation-averaged force remains unaffected by the application on an external flow field. Formally, we anticipate that this will happen in cases where the timescales of the rotational and translational motions are sufficiently separated and the dynamics of the chiral active system can effectively be simulated using point particles interacting via orientation-averaged forces. 
In practice we find that even in cases without a dramatic separation of timescales, the orientation-averaged force can be insensitive to the presence of an applied flow field in practice. We provide numerical evidence for one such example in \secname~\ref{sec:sim_flow}.

\subsection{Distortion modes of pair correlation functions in a flow field} \label{sec:gr_deform}

\begin{figure}[tbp]
	\centering
	\includegraphics[width=0.48\textwidth]{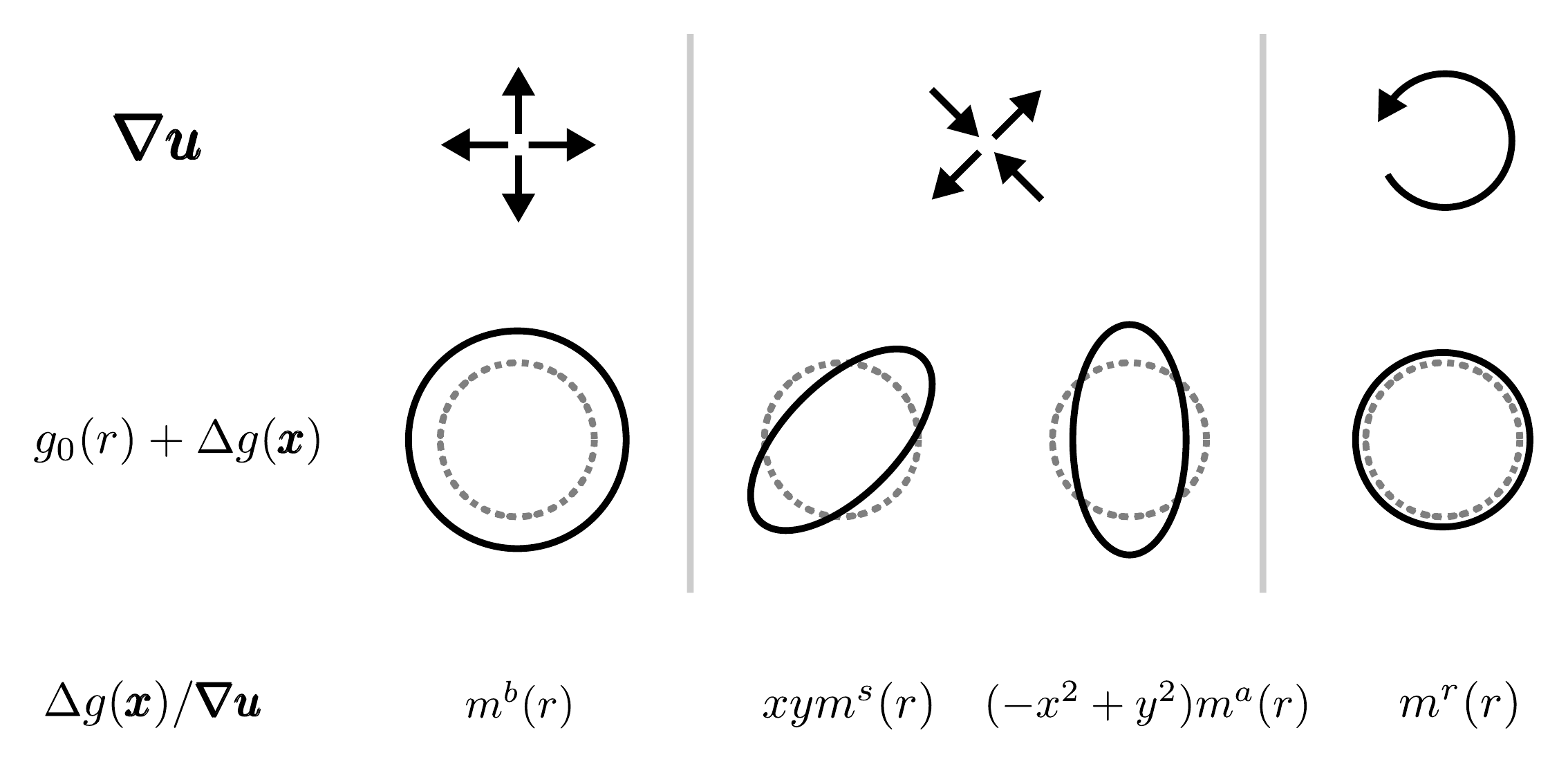}
    \caption{
        Schematic of the four modes of $g(\bm{x})$-distortion induced by a velocity gradient $\nabla \bm{u}$. The first row depicts three types of velocity gradient, dilation, pure shear flow, and vortical flow. The second row shows the corresponding modes of $g(\bm{x})$-distortion. Dotted circle represent the first peak of the undistorted $g(\bm{x})$, and the solid circles or ellipses are schematics of the distorted $g(\bm{x})$. In particular, the pure shear sketched in the middle column can induce two modes of distortion, denoted by $m^s$ and $m^a$. The third row labels these distortion modes according to \eqname~\eqref{eq:gr_dilation}-\eqref{eq:gr_vortical}.
    }
    \label{fig:schem_gr}
\end{figure}

We now derive expressions for the distortion modes of the pair correlation function $g(\bm{x})$ induced by an applied flow field. Our derivation, motivated by work in Ref.~\cite{Kirkwood1949StatisticalMechanical,Avron1998OddViscosity}, relies extensively on symmetry considerations. The symmetry argument in Ref.\cite{Avron1998OddViscosity} is applied at the level of the viscosity tensor. In contrast, our symmetry argument is applied at the level of $g(\bm{x})$, which provides more microscopic details.
Similar symmetry considerations were also used extensively in studies of distortions in $g(\bm{x})$ due to shear flows \cite{Clark1980ObservationCoupling,Hess1980ShearflowinducedDistortion,Hess1982DistortionStructure,Hanley1987ShearinducedAngular}. 
Unlike these early studies which focused mainly on conventional fluids, we focus on terms that can emerge due to chirality and activity. 
The central results of this section show that the distortions of the pair correlation function due to the imposition of a flow field can be captured in terms of four modes as described in \eqname~\eqref{eq:gr_four}-\eqref{eq:gr_vortical} and sketched in \figurename~\ref{fig:schem_gr}.

We begin by considering the pair correlation function in the absence of any imposed flows, $g_0(\bm{x}) = g_0(r)$, and impose perturbative velocity gradients $\partial_ku_l$, where $u_l$ is the $l$'th component of the velocity. We then expand $g(\bm{x})$ to first order in $\partial_ku_l$ and to second order in $x_i$ ($x_1 \equiv x, x_2 \equiv y$) as
\begin{equation} \label{eq:gr_raw}
    g(\bm{x}) = g_0(r) + M_{kl}^{(2)}(r) \partial_ku_l + M_{ijkl}^{(4)}(r) x_ix_j\partial_ku_l + \dots,
\end{equation}
where we have used Einstein's summation convention. 
The coefficients $M(r)$ describe the radial dependence, and factors $x_ix_j$ describe the angular dependence \cite{Hess1982DistortionStructure}. Higher orders of angular dependence, e.g. $x_ix_jx_kx_l$, are omitted here.

Isotropy requires that when the flow field $\bm{\nabla} \bm{u}$ is rotated by an angle $\theta$, the $g(\bm{x})$ is also rotated by $\theta$, but its shape remains unchanged. This constraint of isotropy dramatically reduces the number of allowed variations in \eqname~\ref{eq:gr_raw}.
Now we illustrate this constraint by taking the term $M_{ijkl}^{(4)}(r) x_ix_j\partial_ku_l$ as an example. 
We denote a rotation operation by angle $\theta$ as $T$,
\begin{equation}
    T = \pmqty{
        \cos\theta & -\sin\theta \\
        \sin\theta & \cos\theta
    }.
\end{equation}
Under a rotated flow field $T_{k'k}\partial_k T_{l'l}u_l$, the $M^{(4)}$ term is transformed to $M_{ijkl'}^{(4)}(r) x_ix_jT_{k'k}\partial_k T_{l'l}u_l$. The value of this transformed term at point $T\bm{x}$ should equal the value of the original one at $\bm{x}$, which gives
\begin{equation} \label{eq:m4_equality}
    M_{i'j'k'l'}^{(4)}(r) T_{i'i}x_iT_{j'j}x_jT_{k'k}\partial_kT_{l'l}u_l = M_{ijkl}^{(4)}(r) x_ix_j\partial_ku_l.
\end{equation}
Since \eqname~\eqref{eq:m4_equality} holds for any $x_i, x_j, \partial_ku_l$, we conclude that isotropy requires that $M^{(4)}_{ijkl}$, and from similar arguments $M^{(2)}_{ij}$, satisfy
\begin{align}
    M^{(2)}_{ij} &= M^{(2)}_{i'j'}T_{i'i}T_{j'j}, \label{eq:m2_iso_def}\\
    M^{(4)}_{ijkl} &= M^{(4)}_{i'j'k'l'}T_{i'i}T_{j'j}T_{k'k}T_{l'l}. \label{eq:m4_iso_def}
\end{align}

In addition to isotropy, $M^{(4)}_{ijkl}$ should also satisfy a symmetry requirement that $M^{(4)}_{ijkl}$ is invariant under the exchange of $i$ and $j$, which results from the expression $M_{ijkl}^{(4)}(r) x_ix_j\partial_ku_l$.

The requirements of isotropy and additional symmetry are easier to address if we express $M^{(2,4)}$ in a Pauli-like basis \cite{Avron1998OddViscosity},
\begin{equation}
    \begin{split}
        P^I &= \pmqty{ 1 & 0 \\ 0 & 1}, \quad
        P^X = \pmqty{ 0 & 1 \\ 1 & 0}, \\
        P^Y &= \pmqty{ 0 & 1 \\ -1 & 0}, \quad
        P^Z = \pmqty{ 1 & 0 \\ 0 & -1}.
    \end{split}    
\end{equation}
Using the Pauli-like basis, $M^{(2,4)}$ are expanded as
\begin{align}
    M^{(2)}_{ij} &= m^{(2)a} P^a_{ij}, a=I,X,Y,Z \\
    M^{(4)}_{ijkl} &= m^{(4)ab} P^a_{ij} P^b_{kl}, a=I,X,Z, b=I,X,Y,Z.
\end{align}

\eqname~\eqref{eq:m2_iso_def}-\eqref{eq:m4_iso_def} now read
\begin{gather}
    (T_{i'i}T_{j'j} - \delta_{i'i}\delta_{j'j}) P^a_{i'j'} m^{(2)a} = 0, \\
    (T_{i'i}T_{j'j}T_{k'k}T_{l'l} - \delta_{i'i}\delta_{j'j}\delta_{k'k}\delta_{l'l}) P^a_{i'j'}P^b_{k'l'} m^{(4)ab} = 0.
\end{gather}

Solving the above linear equations for $m^{(2)a}, m^{(4)ab}$, we get the allowed forms of $M^{(2,4)}$,
\begin{align} \label{eq:m_iso_expr}
    M^{(2)}_{kl} &= m^{(2)s} P^I_{kl} + m^{(2)r} P^Y_{kl}, \\
    M^{(4)}_{ijkl} &= m^{(4)I} P^I_{ij}P^I_{kl} +
        m^{(4)s} (P^X_{ij}P^X_{kl} + P^Z_{ij}P^Z_{kl}) + \\
        & m^{(4)a} (P^X_{ij}P^Z_{kl} - P^Z_{ij}P^X_{kl}) +
        m^{(4)r} P^I_{ij}P^Y_{kl},
\end{align}
where $m^{(2)s}, m^{(2)r}$, etc. are arbitrary functions of $r$. 

Plugging the expressions for $M^{(2,4)}$ into the expansion of $g(\bm{x})$ \eqname~\eqref{eq:gr_raw} and grouping terms, we obtain the allowed distortions of $g(\bm{x})$.
These distortions are responses to the strain rate $\nu_{kl}$ and vorticity $\omega$, defined as
\begin{align}
    \nu_{kl} &= \frac{1}{2}(\partial_ku_l + \partial_lu_k), \\
    \omega &= \frac{1}{2}(\partial_xu_y - \partial_yu_x).
\end{align}
The distorted $g(\bm{x})$ consists of, in addition to the unperturbed $g_0(r)$, four distortion modes,
\begin{equation} \label{eq:gr_four}
    g(\bm{x}) = g_0(r) + g^b(\bm{x}) + g^s(\bm{x}) + g^a(\bm{x}) + g^r(\bm{x}).
\end{equation} 
Each mode reads
\begin{align}
    g^b(\bm{x}) &= m^b(r) (\nu_{xx}+\nu_{yy}), \label{eq:gr_dilation}\\
    g^s(\bm{x}) &= m^s(r) \pmqty{x & y} \pmqty{
        \nu_{xx} - \nu_{yy} & 2\nu_{xy} \\
        2\nu_{yx} & \nu_{yy} - \nu_{xx}
    } \pmqty{x \\ y}, \label{eq:gr_shear_normal} \\
    g^a(\bm{x}) &= m^a(r) \pmqty{x & y} \pmqty{
        - (\nu_{xy}+\nu_{yx}) & \nu_{xx}-\nu_{yy}\\
        \nu_{xx}-\nu_{yy} & \nu_{xy}+\nu_{yx}
    } \pmqty{x \\ y}, \label{eq:gr_shear_anomalous}\\
    g^r(\bm{x}) &= m^r(r) \omega, \label{eq:gr_vortical}
\end{align}
where $m^b,m^s,m^a,m^r$ are undetermined functions that depend only on the scalar $r$. The original functions like $m^{(2)s}$ are grouped into these new functions through, for example, $m^b = m^{(2)s} + m^{(4)I} r^2$.
Determining the functions $m^{b,s,a,r}(r)$ requires theories starting from a given microscopic equation of motion or simulations of specific systems. 
To keep our discussion general, we retain these undetermined functions $m^{b,s,a,r}(r)$.

The modes of distortion in \eqname~\eqref{eq:gr_dilation}-Eq.~\ref{eq:gr_vortical} can be interpreted as follows. The term $g^b$ describes the distortion of the pair correlation function induced by dilation, while the terms $g^s$ and $g^a$ describe distortions induced by shear strain, and the term $g^r$ describes a distortion induced by vortical flows.
The four modes of distortion are illustrated graphically in \figurename~\ref{fig:schem_gr}. 
For ordinary liquids with parity symmetry, the pair correlation function computed in the presence of a pure shear should be invariant under the exchange of $x$ and $y$. Hence the distortion in the pair correlation function will not have a mode corresponding to the term $g^a$. Similar considerations in the presence of vortical flows allow us to rule out the distortion mode $g^r$. Thus from ordinary liquids with parity symmetry, the only allowed modes of distortion are $g^b$ and $g^s$. 
For chiral fluids, however, there are no \textit{a priori} constraints on these modes of distortion, so all four distortions $g^b, g^s, g^a, g^r$ are possible.
It should be noted that these constraints are based on the ansatz of $g(\bm{x})$, \eqname~\eqref{eq:gr_raw}, where $\partial_ku_l$ is expanded to its first order and $x_i$ is expanded to its second order. These constraints can break even for ordinary fluids when higher-order contributions are considered \cite{Hess1980ShearflowinducedDistortion,Hess1982DistortionStructure,Hanley1987ShearinducedAngular}. 

\subsection{Anomalous transport coefficients} \label{sec:transport}

Using \eqname~\eqref{eq:gr_dilation}-\eqref{eq:gr_vortical} and the expression for the stress tensor in \eqname~\eqref{eq:stress_IK}, we can now extract the transport coefficients. 
The final results, displayed in \eqname~\eqref{eq:stress_terms}-\eqref{eq:stress_odd_bulk}, are combinations of the averaged forces $\bar{F}_{\parallel /\perp}$ and the modes in \eqname~\eqref{eq:gr_four}.

For ease of notations, we write the parallel(perpendicular) force as the gradient of a formal parallel(perpendicular) potential, $V'_\parallel(r)$($V'_\perp(r)$).
\begin{equation}
    \bar{\bm{F}}(\bm{x}) = - \frac{V'_\parallel(r)}{r} \pmqty{x \\ y} - \frac{V'_\perp(r)}{r} \pmqty{-y \\ x}.
\end{equation}
We also introduce a notation for the average
\begin{align}
    \expval{A}_{\parallel/\perp} = \int_0^\infty \dd{r} \frac{V'_{\parallel/\perp}(r)\rho^2}{2} \pi r^2 A(r).
\end{align}

After straightforward calculations, the final stress tensor can be written as,
\begin{equation} \label{eq:stress_terms}
    \begin{split}
        \bm{\sigma} 
        =& \bm{\sigma}_{e,0} + \bm{\sigma}_{e,r} + \bm{\sigma}_{e,b} + \bm{\sigma}_{e,s} + \bm{\sigma}_{e,a} + \\
        &\bm{\sigma}_{o,0} + \bm{\sigma}_{o,r} + \bm{\sigma}_{o,b}
    \end{split}
\end{equation}
where the subscript $e(o)$ indicate that the term is symmetric(antisymmetric). Terms in \eqname~\ref{eq:stress_terms} are defined as follows:

The component $\bm{\sigma}_{e,0}$ is an ordinary pressure-like term. Using $I$ to denote the identity matrix, $\bm{\sigma}_{e,0}$ reads
\begin{equation} \label{eq:stress_pressure}
    \bm{\sigma}_{e,0} = \expval{g_0}_\parallel I.
\end{equation}

The component $\bm{\sigma}_{e,b}$ describes a symmetric stress induced by dilation, from which bulk viscosity is extracted as $\expval{m^b}_\parallel$.
\begin{equation} \label{eq:stress_bulk_vis}
    \bm{\sigma}_{e,b} = \expval{m^b}_\parallel (\nu_{xx}+\nu_{yy}) I.
\end{equation}

The component $\bm{\sigma}_{e,s}$ describes the stress from shear viscosity $\eta_{\rm s}\equiv \expval{r^2m^s/2}_\parallel - \expval{r^2m^a/2}_\perp$,
\begin{equation} \label{eq:stress_shear_vis}
    \bm{\sigma}_{e,s} = \eta_{\rm s}
    \pmqty{
        \nu_{xx}-\nu_{yy} & 2\nu_{xy} \\
        2\nu_{yx} & \nu_{yy}-\nu_{xx}
    }.
\end{equation}

The component $\bm{\sigma}_{e,a}$ describes the stress from odd viscosity $\eta_a\equiv \expval{r^2m^s/2}_\perp + \expval{r^2m^a/2}_\parallel$,
\begin{equation} \label{eq:stress_odd_vis}
    \bm{\sigma}_{e,a} = \eta_a
    \pmqty{
        -(\nu_{xy}+\nu_{yx}) & \nu_{xx}-\nu_{yy} \\
        \nu_{xx}-\nu_{yy} & \nu_{xy}+\nu_{yx}
    }.
\end{equation}
The odd viscosity is so-called because its tensor form $\eta_{ijkl}$, which relates the even part of stress and strain rate through $\sigma_{e,ij} = \eta_{ijkl}\nu_{kl}$, is antisymmetric $\eta_{ijkl}=-\eta_{klij}$ \cite{Avron1998OddViscosity}. For isotropic systems, the antisymmetric tensor has only one independent component \cite{Avron1998OddViscosity}, which we simply call the odd viscosity coefficient. This odd viscosity, unlike the conventional shear viscosity, does not lead to dissipation \cite{Banerjee2017OddViscosity}.

The terms $\bm{\sigma}_{o,0}$ and $\bm{\sigma}_{o,r}$ describe two antisymmetric components of the stress tensor, one is static and one reflects a response to vortical flows,
\begin{equation} \label{eq:stress_odd}
    \bm{\sigma}_{o,0} + \bm{\sigma}_{o,r}
    = (\expval{g_0}_\perp + \expval{m^r}_\perp\omega)\pmqty{
        0 & -1 \\
        1 & 0
    }.
\end{equation}

The component $\bm{\sigma}_{e,r}$ describes a diagonal stress that responds to vortical flows,
\begin{equation} \label{eq:stress_diag_rot}
    \bm{\sigma}_{e,r} = \expval{m^r}_\parallel \omega I.
\end{equation}

Finally, the component $\bm{\sigma}_{o,b}$ describes another antisymmetric part of the stress tensor, which responds to dilations,
\begin{equation} \label{eq:stress_odd_bulk}
    \bm{\sigma}_{o,b} = \expval{m^b}_\perp (\nu_{xx}+\nu_{yy}) \pmqty{
        0 & -1 \\
        1 & 0
    }.
\end{equation}

Note that terms containing $m^b,m^s$ or $\expval{\cdot}_\parallel$ exist even in regular achiral liquids, whereas terms containing $m^a,m^r$ or $\expval{\cdot}_\perp$ are specific to chiral fluids.
For ordinary achiral liquids, $m^a,m^r$ and $\expval{\cdot}_\perp$ vanish, and the terms in the stress expression reduce to the usual pressure, bulk viscosity, and shear viscosity.
For chiral liquids, however, all these terms are possible. 
In particular, the odd viscosity coefficient $\eta_a$ has two contributions, one from the perpendicular force acting on the ordinary distortion of $g(\bm{x})$, and the other from the ordinary parallel force acting on the anomalous distortion of $g(\bm{x})$.
We also see that the shear viscosity gets modified by $\expval{r^2m^a/2}_\perp$, which comes from the perpendicular force acting on the anomalous distortion of $g(\bm{x})$.

We briefly compare our components of stress tensor with the literature. Terms \eqname~\eqref{eq:stress_pressure}-\eqref{eq:stress_odd} agree well with Ref.~\cite{Banerjee2017OddViscosity}, where transport properties were derived starting from hydrodynamic equations.
We also obtain two additional terms, \eqname~\eqref{eq:stress_diag_rot},\eqref{eq:stress_odd_bulk}. Including these two terms, we found six coefficients that relate stress tensor and velocity gradients. This result is consistent with Ref.\cite{Scheibner2019OddElasticity}, which derived a generalized viscosity tensor that satisfies isotropy condition.

\section{Numerical simulations of a model active rotor system} \label{sec:simulation}

In this section, we present results from numerical simulation of a model active rotor system. 
First, we perform numerical simulations in the absence of any imposed gradients in the velocity fields (\secname~\ref{sec:sim_force}), and show that the orientation-averaged inter-rotor forces can contain a perpendicular component.
Next, from the simulation of rotors under an imposed shear flow (\secname~\ref{sec:sim_flow}), we show the orientation-averaged intermolecular forces do not change under this flow field for our model rotor system. We also show in \secname~\ref{sec:gr_expt} that the distortion of pair correlation function to linear order in $\partial_ku_l$ is well described by the modes \eqname~\eqref{eq:gr_shear_normal}-\eqref{eq:gr_vortical}. Thus, the above described procedure for extracting transport coefficients can be applied to our system. Consequently in \secname~\ref{sec:num_est}, we obtain numerical estimates of the various transport coefficients using the averaged force and $g(\bm{x})$-distortions extracted from the simulations.

\subsection{Perpendicular component of the orientation-averaged force} \label{sec:sim_force}

\begin{figure}[tbp]
	\centering
	\includegraphics[width=0.48\textwidth]{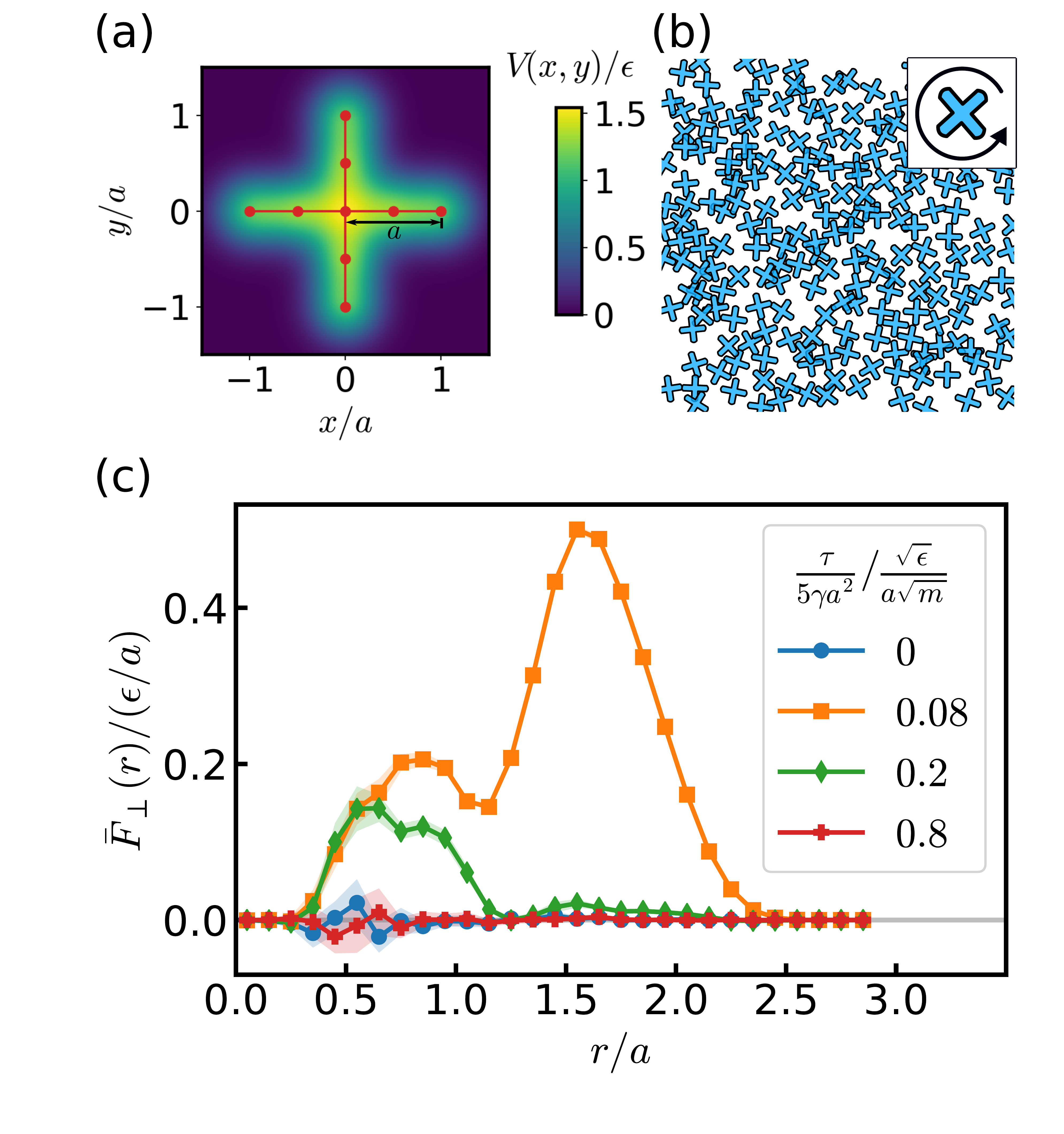}
    \caption{
        Transverse components in the orientation-averaged intermolecular forces.
        (a) The inter-rotor potential used in our numerical simulations. Each rotor is simulated by nine beads, which are labeled as red dots. Beads belonging to different rotors interact via a Gaussian potential $V(r) = \epsilon e^{-r^2/2\sigma^2}, \sigma=0.25a$. The figure plots the superposition of the Gaussian potential from the nine beads.
        (b) A snapshot of a section of the simulation system. Each rotor is driven in a counter-clockwise direction.
        (c) Perpendicular component of the averaged inter-rotor forces at different torques. Shaded region around each curve represent $95\%$ confidence interval. $\tau$ is the external torque, and $\tau/5\gamma a^2$ is the average angular velocity of rotors under the external torque when the rotors are non-interacting.
    }
    \label{fig:force}
\end{figure}

The rotors in our model chiral active liquid are constructed using nine beads held fixed relative to one another in a set geometry. Beads belonging to different rotors interact via a Gaussian potential $V(r) = \epsilon e^{-r^2/2\sigma^2}, \sigma=0.25a$, where $\epsilon$ sets the energy scale, and $a$ denotes the length-scale of each rotor. In \figurename~\ref{fig:force} we plot the superposition of the Gaussian potential from the nine beads in a rotor.
The dynamics of rotors are simulated using underdamped Langevin dynamics. 
The molecular dynamics simulations were performed using LAMMPS package \cite{Plimpton1995FastParallel} with Moltemplate toolkit \cite{Jewett2013MoltemplateCoarseGrained} and custom code. The friction of Langevin dynamics $\gamma$ is set to $\gamma = 0.5 \sqrt{m\epsilon}/a$, where $m$ is the mass of each bead, and the temperature is $\epsilon/k_B$.
Timestep is set to $0.0004 a\sqrt{m/\epsilon}$. The numerical results described below were obtained from simulations performed with $N=900$ rotors in a simulation box with dimensions $60a\times 60a$ (\figurename~\ref{fig:force}b). The simulations were performed with periodic boundary conditions. 
A constant counter-clockwise torque $\tau$ is applied to each rotor. When applied to a single rotor, this torque would cause the rotor to rotate with an averaged angular velocity $\omega_0 = \tau/5\gamma a^2$. We performed simulations with multiple values of applied torque, $\tau/\epsilon = 0, 0.2, 0.5, 2$. 
For each value of torque, we collected data from $10$ steady-state trajectories, each with length $2000 a\sqrt{m/\epsilon}$.

The perpendicular component of the orientation-averaged forces computed from simulations are plotted in \figurename~\ref{fig:force}c. 
For this particular rotor model, significant $\bar{F}_\perp(\bm{x})$ emerges at intermediate torques ($\omega_0 / \sqrt{\epsilon/ma^2} = 0.08, 0.2$) before decreasing to vanishingly small values at large torques ($\omega_0 / \sqrt{\epsilon/ma^2} = 0.8$).

\subsection{Simulation of rotors under a flow field} \label{sec:sim_flow}

\begin{figure}[tbp]
	\centering
	\includegraphics[width=0.48\textwidth]{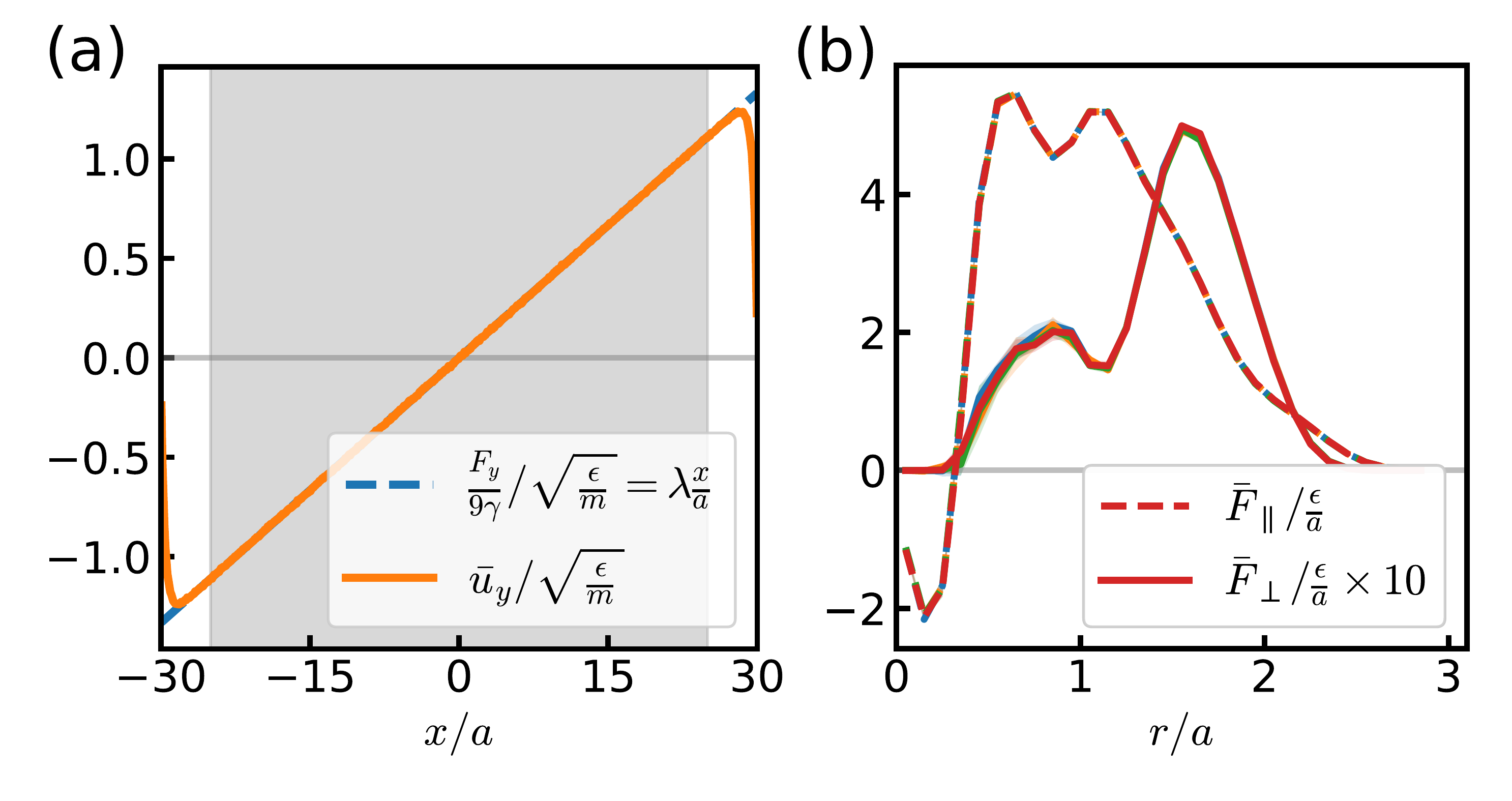}
    \caption{
        Averaged parallel and perpendicular force from simulated rotor systems under a flow field.
        (a) Profile of applied external force $F_y$ and the measured averaged velocity $u_y$ in $y$-direction. Data is shown for $\lambda = 0.0444$. Due to periodic boundary condition, $u_y$ close to the boundary $|x|=30a$ decreases to zero. Calculation of inter-rotor forces only include rotors whose center of masses are inside the shaded region ($|x|<25a$), where the averaged velocity is linear in $x$.
        (b) $\bar{F}_\parallel(r)$ and $\bar{F}_\perp(r)$ for $\lambda = 0, 0.0222, 0.0444, -0.0444$. $\bar{F}_{\parallel/\perp}(r)$ at different $\lambda$'s basically overlap, which shows that small applied flow fields do not induce significant changes in $\bar{F}_{\parallel/\perp}(r)$. To avoid overcrowding the plot, legends for colors corresponding to different $\lambda$'s are not shown. Shaded region around each curve represent $95\%$ confidence interval, which is small compared with the line width.
    }
    \label{fig:flow}
\end{figure}

\begin{figure}[tbp]
	\centering
	\includegraphics[width=0.48\textwidth]{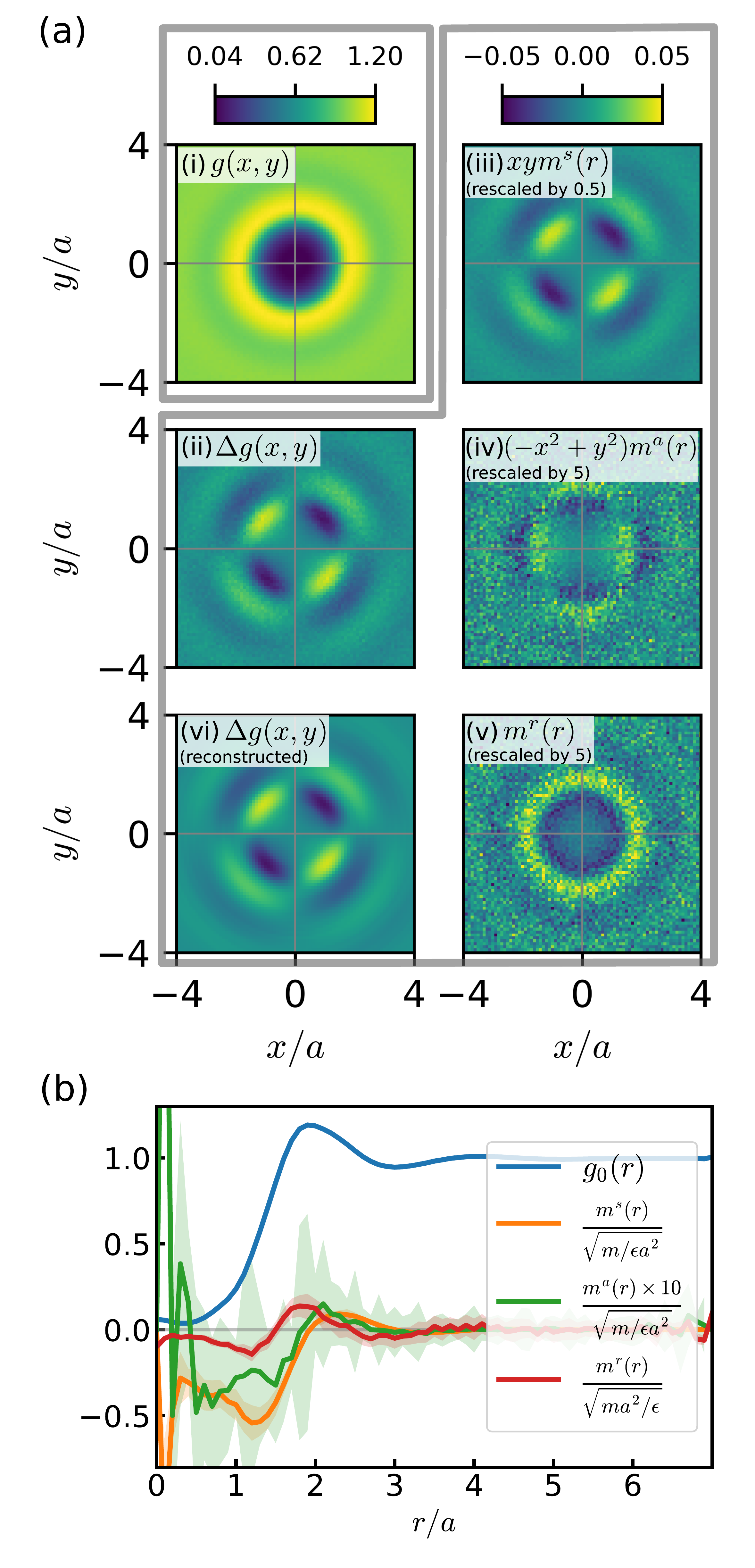}
    \caption{
        Distortion of the pair correlation function induced by a simple shear flow field.
        (a) 
        $g(x, y)$ under a simple shear flow $\partial_xu_y = 0.0444 \sqrt{\epsilon/ma^2}$ is distorted to an elliptical shape (i).
        Apparently, the distortion $\Delta g(x, y)$ (ii) is slightly rotated counter-clockwisely compared with that in ordinary liquids. 
        (iii) plots the mode $(\Delta g(x, y) - \Delta g(x, -y))$, which is proportional to $xym^s(r)$, and is labeled by $xym^s(r)$ for simplicity. The data is rescaled by a factor of $0.5$ to accommodate the scale of the colorbar.
        The other two modes, $(\Delta g(x, y) - \Delta g(y, x)) \propto (-x^2+y^2)m^a(r)$ and $(\Delta g(x, -y) + \Delta g(y, x)) \propto m^r(r)$ are plotted in (iv) and (v), respectively.
        The $\Delta g(x, y)$ reconstructed from the fitted $m^{s,a,r}(r)$ (vi) agrees well with $\Delta g(x, y)$ directly computed from simulation (iii).
        (b) $m^{s,a,r}(r)$ fitted from the data in (a). Shaded region around each curve represent the standard deviation of residues at each $r$.
    }
    \label{fig:gr_distortion}
\end{figure}

The simulation setup of this subsection is the same as the one in \secname~\ref{sec:sim_force}, except that an additional external force $F_y / \sqrt{\epsilon/m} = 9\gamma \lambda x / a$ is applied to the center of mass of each rotor. This applied external force creates a simple shear in the region around $x=0$. Rotors' averaged velocity in $y$-direction satisfies $u_y / \sqrt{\epsilon/m} = F_y / 9\gamma = \lambda x / a$ (\figurename~\ref{fig:flow}a, except for rotors close to the left and right boundaries).
The simulation parameters are identical to the ones used in \secname~\ref{sec:sim_force}, except that the torque applied on each rotor is set to $\tau=0.2 \epsilon$, and we vary the value of $\lambda$ for different set of simulations.
We collected data from steady-state trajectories of length $2000 a\sqrt{m/\epsilon}$. We computed the averaged inter-rotor forces using $10$ trajectories while we used $100$ trajectories to compute the $g(\bm{x})$-distortions.
For the purposes of these computations we only include rotors in the middle region of the simulation box in \figurename~\ref{fig:flow}(a), where the velocity gradient $\partial_x u_y = \lambda \sqrt{\epsilon/ma^2}$ is constant. In more detail, the $x$-coordinate of the center of mass of each rotor considered for our analysis is within $[-25a, 25a]$ for computing inter-rotor forces, and is within $[-20a, 20a]$ for computing $g(\bm{x})$.

The averaged inter-rotor forces computed from simulations are described in \figurename~\ref{fig:flow}(b). We see that there is no significant change in the averaged intermolecular forces for various values of $\lambda$. Hence, the technique outlined in Sec.~\ref{sec:extract} can be used to obtain estimates of the various transport coefficients.

\subsubsection{Extracting deformations in the pair correlation functions due to imposed flow: results from numerical simulations} \label{sec:gr_expt}

The distortion of pair correlation function induced by a flow field can be extracted as follows. The simple shear flow we imposed is a superposition of pure shear and vortical flows, $\partial_x u_y = \nu_{xy} + \omega$, and according to \eqname~\eqref{eq:gr_four}-\eqref{eq:gr_vortical}, the ansatz of $g(\bm{x})$-distortion reads
\begin{align}
    \Delta g(\bm{x}) &= \partial_x u_y[2xym^s(r) + (-x^2+y^2)m^a(r) + \frac{1}{2}m^r(r)], \label{eq:flow_gr_three} \\
    \Delta g(\bm{x}) &= \frac{1}{2}(g(\bm{x}; \partial_xu_y) - g(\bm{x}; -\partial u_y)). \label{eq:flow_delta_gr}
\end{align}
Here $\Delta g(\bm{x})$ is calculated using $g(\bm{x})$'s under both $\partial_xu_y$ and $-\partial_xu_y$ in order to eliminate possible quadratic terms ($\propto (\partial_xu_y)^2$). At the flow field magnitude we used in simulation ($\lambda = 0.0444$), these quadratic terms do contribute to the $g(\bm{x})$-distortion. Smaller flow field magnitudes are not adopted because they produce poor signal-to-noise ratios, and in fact, they are not required because, as we will show below, the linear distortions can already be extracted from $\Delta g(\bm{x})$.
Utilizing symmetries of \eqname~\eqref{eq:flow_gr_three}, the distortion modes can be computed as
\begin{align}
    m^s(r) &= \frac{1}{4xy\partial_xu_y} (\Delta g(x, y) - \Delta g(x, -y)), \label{eq:flow_ms} \\
    m^a(r) &= \frac{1}{2(-x^2 + y^2)\partial_xu_y} (\Delta g(x, y) - \Delta g(y, x)), \label{eq:flow_ma} \\
    m^r(r) &= \frac{1}{\partial_xu_y} (\Delta g(x, -y) + \Delta g(y, x)). \label{eq:flow_mr}
\end{align}
Numerically, division by $xy$ or $(-x^2+y^2)$ is not favored because their values can be zero. Our actual procedure was to first compute $\Delta g(x, y) - \Delta g(x, -y)$, then find $m^s(r)$ such that $xym^s(r)$ best fits this computed data. 

Simulated $g(\bm{x})$-distortion and its modes are shown in \figurename~\ref{fig:gr_distortion}. We see that $\Delta g(x, y)$ reconstructed from fitted $m^{s,a,r}(r)$ agree well with $\Delta g(x, y)$ directly computed from simulated data (\figurename~\ref{fig:gr_distortion}(a-ii,vi)). This agreement justifies the ansatz of $\Delta g(\bm{x})$ in \eqname~\eqref{eq:flow_gr_three}, which means that we can ignore higher order terms like $x_{i1}x_{i2}x_{i3}x_{i4} \partial_ku_l$ (as discussed in Ref.\cite{Hanley1987ShearinducedAngular}) and $(\partial_ku_l)^3$.
Notably, the simulation results show that the anomalous $g(\bm{x})$-distortions described by $(-x^2+y^2)m^a(r)$ and $m^r(r)$ do exist in our parity symmetry-breaking rotor system.

\subsubsection{Numerical estimates of transport coefficients}\label{sec:num_est}
With the averaged inter-rotor forces $\bar{F}_{\parallel/\perp}(r)$ and the modes of $g(\bm{x})$-distortion $m^{s,a,r}(r)$, we can compute the transport coefficients according to \eqname~\eqref{eq:stress_pressure}-\eqref{eq:stress_diag_rot}. For our model rotor system, the two contributions to the odd viscosity dictated by \eqname~\eqref{eq:stress_odd_vis} are
\begin{align}
    \expval{m^sr^2/2}_\perp &= (0.0220\pm 0.0005) \sqrt{m\epsilon/a^2}, \\
    \expval{m^ar^2/2}_\parallel &= (0.010\pm 0.003) \sqrt{m\epsilon/a^2},
\end{align}
which produce an odd viscosity of $(0.032\pm 0.003) \sqrt{m\epsilon/a^2}$.
Note that the two terms contributing to odd viscosity are on the same order of magnitude, so we cannot simply ignore either one. For comparison, the two contributions to the shear viscosity dictated by \eqname~\eqref{eq:stress_shear_vis} are 
\begin{align}
    \expval{m^sr^2/2}_\parallel &= (0.210\pm 0.004) \sqrt{m\epsilon/a^2}, \\
    \expval{m^ar^2/2}_\perp &= (0.0009\pm 0.0005) \sqrt{m\epsilon/a^2},
\end{align}
which produce a shear viscosity of $(0.209\pm 0.004) \sqrt{m\epsilon/a^2}$. Compared to $\expval{m^sr^2/2}_\parallel$, the term $\expval{m^ar^2/2}_\perp$ is negligible.
The ratio of odd viscosity to shear viscosity is $0.15\pm 0.01$. Finally another anomalous transport property, the antisymmetric stress (\eqname~\eqref{eq:stress_odd}) for our rotor model evaluates to
\begin{align}
    \expval{g_0}_\perp &= (-0.09327\pm 0.00007) \epsilon/a^2,\\
    \expval{m^r}_\perp &= (-0.0052 \pm 0.0004) \sqrt{m\epsilon/a^2}.
\end{align}

\section{Conclusions} \label{sec:conclusion}
In conclusion, we have provided a mechanism for how anomalous transport coefficients can emerge in chiral active fluids. 
The central results of our work are formulated in \eqname~\eqref{eq:stress_terms}-\eqref{eq:stress_odd_bulk}.
In particular, we introduced an orientation-averaged intermolecular force, derived four allowed distortion modes of the pair correlation function in a flow field, and showed how a perpendicular component in the orientation-averaged force and the anomalous distortions of $g(\bm{x})$ combine to produce anomalous transport coefficients such as odd viscosity.
By decomposing the contribution to transport coefficients in terms of averaged forces and distortions of $g(\bm{x})$, we have provided a microscopic perspective to understand these transport properties. 
In future work, we expect to investigate computationally or experimentally how the molecular structure and intermolecular interactions in chiral active liquids affect $\bar{\bm{F}}(\bm{x})$ and $g(\bm{x})$-distortion, thus determining the transport properties.

\begin{acknowledgments}
    This work was primarily supported by NSF DMR-MRSEC 1420709. SV acknowledges support from the Sloan Foundation, startup funds from the University of Chicago and support from the National Science Foundation under award number DMR-1848306. MF and MH acknowledge support from the University of Chicago MRSEC through a Kadanoff-Rice postdoctoral fellowship.
\end{acknowledgments}

%

\end{document}